\documentclass[3p,times,procedia]{elsarticle}
\usepackage{nupha_ecrc}

\volume{00}

\firstpage{1}

\journalname{Nuclear Physics A}

\runauth{A. Sch\"afer et al.}

\jid{nupha}

\jnltitlelogo{Nuclear Physics A}

\usepackage{amssymb}

\usepackage[figuresright]{rotating}

\usepackage{booktabs}
\usepackage{amsmath}

\begin{document}

\begin{frontmatter}



\dochead{XXVIIIth International Conference on Ultrarelativistic Nucleus-Nucleus Collisions\\ (Quark Matter 2019)}

\title{A Non-Equilibrium Approach to Photon Emission from the Late Stages of Relativistic Heavy-Ion Collisions }


\author[1,2]{Anna Sch\"afer}
\ead{aschaefer@fias.uni-frankfurt.de}
\author[2]{Juan M. Torres-Rincon}
\author[3]{Charles Gale}
\author[4,2,1]{Hannah Elfner}

\address[1]{Frankfurt Institute for Advanced Studies, Ruth-Moufang-Strasse 1, 60438 Frankfurt am Main, Germany}
\address[2]{Institute for Theoretical Physics, Goethe University, Max-von-Laue-Strasse 1, 60438 Frankfurt am Main, Germany}
\address[3]{Department of Physics, McGill University, 3600 University Street, Montreal, QC, H3A 2T8, Canada}
\address[4]{GSI Helmholtzzentrum f\"ur Schwerionenforschung, Planckstrasse 1, 64291 Darmstadt, Germany}

\begin{abstract}
Cross sections for photon production in hadronic scattering processes have been calculated according to an effective chiral field theory.
For $\pi + \rho \to \pi + \gamma$ and $\pi + \pi \to \rho + \gamma$ processes, these cross sections have been implemented into a novel hadronic transport approach (SMASH), which is suitable for collisions at low and intermediate energies.
The implementation is verified by systematically comparing the thermal photon rate to theoretical expectations. The photon rates we obtain are compared to previous works, where scattering processes mediated by $\omega$ mesons are found to contribute significantly to the total photon production.
Finally, the impact of considering the finite width of the $\rho$ meson is investigated, and a significant enhancement of photon production in the low-energy region is observed. This work is the first step towards a consistent treatment of photon emission in hybrid hydrodynamics+transport approaches. The quantification of the importance of the hadronic stage for the resolution of the \textit{direct photon flow puzzle} is a next step and can be applied to identify equilibrium and non-equilibrium effects in the hadronic afterburner.
\end{abstract}


\end{frontmatter}


\section{Introduction}
\label{Intro}
Photons are direct and unique probes in heavy-ion collisions.
They escape the fireball unaffected owing to their purely electromagnetic coupling and are produced in all stages of the collision.
Hence, they draw a time-integrated picture of the entire evolution, carrying properties from the medium to the detector.
It is therefore essential to fully understand the production mechanisms and properties of photons in each stage of the collision.
This work focuses on the late stages, where the fireball is believed to have expanded and cooled down sufficiently to find quarks and gluons confined again into hadrons.
For this purpose, a non-equilibrium approach for photon emission from hadronic interactions is presented.

\section{Model Description}
The photon production framework introduced in the following is based on the SMASH (Simulating Many Accelerated Strongly-interacting Hadrons) hadronic transport approach \cite{Weil:2016zrk, SMASH-DOI}.
It is designed for the description of heavy-ion collisions at low and intermediate energies.
Furthermore it was successfully applied as an hadronic afterburner \cite{Ryu:2019atv, JF_QM2020} which allows for a non-equilibrium study of the late stages of relativistic heavy-ion collisions.
In SMASH, photon production in binary, hadronic scattering processes is implemented based on an effective chiral field theory with mesonic degrees of freedom \cite{Schafer:2019edr}.
Among those are pseudoscalar mesons, vector mesons, axial vector mesons and the photon.
The corresponding Lagrangian reads
\begin{align}
  	\mathcal{L} = & \ \dfrac{1}{8} \ F_{\pi}^2 \ Tr\left(D_{\mu}UD^{\mu}
  	U^{\dagger}\right) \ + \ \dfrac{1}{8} \ F_{\pi}^2 \  Tr\left(M\left(U
  	+U^{\dagger}-2\right)\right) - \ \dfrac{1}{2} \ Tr \left(F_{\mu\nu}
  	^{L}F^{L \mu\nu} \ + \ F_{\mu\nu}^{R} \ F^{R \mu\nu} \right) \notag
    	+ \\&
   	 \ m_0^2 \ Tr\left(A_{\mu}^LA^{L\mu} + A_{\mu}^RA^{R\mu} \right) \
  	+ \ \gamma \ Tr \left(F_{\mu\nu}^{L}U F^{R \mu\nu} U^{\dagger}
  	\right) \ \notag  - \ i \xi \ Tr \left(D_{\mu}UD_{\nu}U^{\dagger}
  	F^{L\mu\nu} + D_{\mu}U^{\dagger}D_{\nu}UF^{R\mu\nu} \right) - \\&
  	\ \dfrac{2em_V^2}{\tilde{g}} B_{\mu} \ Tr \Bigl(Q
  	\tilde{V}^{\mu}\Bigr) \ - \ \dfrac{1}{4}
  	\left(\partial_{\mu} B^{\nu} - \partial_{\nu} B^{\mu}\right)^2 + \
  	\dfrac{2 e^2 m_0^2}{g_0^2} B_{\mu}B^{\mu} Tr\left(Q^2\right)
  	+ \ g_{VV\phi} \ \varepsilon_{\mu\nu\alpha\beta} \
  	Tr \Big[\partial^{\mu} V^{\nu} \partial^{\alpha} V^{\beta} \phi \Big]
  	\label{Lagrangian}
  \end{align}

and contains the couplings and interaction of the aforementioned particles. For further details about the underlying theoretical framework the reader is referred to \cite{Turbide:2003si}. \\
Based on the properties enclosed in Equation~(\ref{Lagrangian}), it is possible to derive the Feynman rules, matrix elements and cross sections for the photon producing scattering processes.
We limit our explorations to processes of the kind $\ \pi + \pi \to \rho + \gamma \ $ and $\ \pi + \rho \to \pi + \gamma\ $ as those provide the leading contributions.
The specific scattering processes considered are collected in Table \ref{channel_table}.
As in \cite{Turbide:2003si}, they are grouped into processes mediated by ($\pi$, $\rho$, $a_1$) mesons (left) and those mediated by the $\omega$ meson (right).
It is evident, that processes (e) and (g) as well as (d) and (h) differ only with respect to their mediating particles but have identical initial and final states.
Those processes are first treated separately, to allow for a straight-forward validation of the presented framework.
Once validated, these processes are incoherently added while accounting for their respective form factors, as described in \cite{Schafer:2019edr}.\\
It should be noted that within the afore presented framework, the $\rho$ meson is assumed to be a stable particle with vanishing width.
It is however known that in reality, $\Gamma_\rho$ = 0.149 GeV \cite{Agashe:2014kda}, such that a more realistic description is necessary.
An attempt is made to extend the presented framework to also describe broad $\rho$ mesons.
The SMASH resonance treatment is employed and the $\rho$ meson mass is sampled from a Breit-Wigner distribution.
The initial or final state $\rho$ masses enter the computation of the photon cross sections directly, higher-order corrections are however not introduced. The disagreement between the $\rho$ meson mass in the loops and the in-/outgoing $\rho$ meson masses introduces a systematic error which is found to be smaller than 11\% in the configurations tested. Further details are explained in \cite{Schafer:2019edr}.

\begin{table}
  \centering
  \begin{tabular}{ c c}
    \toprule
    Mediated by ($\pi$, $\rho$, $a_1$) mesons & $\ \ $ Mediated by $\omega$ meson $\ \ $ \\[+0.1cm]
    \hline
     & \\[-0.2cm]
    $\pi^\pm + \pi^\mp \to \rho^0 + \gamma \quad $  (a) & $\pi^0 + \rho^0 \to \pi^0 + \gamma \quad $  (f) \\[+0.1cm]
    $\pi^\pm + \pi^0 \to \rho^\pm + \gamma \quad $  (b) & $\pi^\pm + \rho^\mp \to \pi^0 + \gamma \quad $  (g) \\[+0.1cm]
    $\pi^\pm + \rho^0 \to \pi^\pm + \gamma \quad $  (c) & $\pi^0 + \rho^\pm \to \pi^\pm + \gamma \quad $  (h) \\[+0.1cm]
    $\pi^0 + \rho^\pm \to \pi^\pm + \gamma \quad $  (d) & \\[+0.1cm]
    $\pi^\pm + \rho^\mp \to \pi^0 + \gamma \quad $  (e) & \\[+0.1cm]
    \bottomrule
  \end{tabular}
  \caption{Photon production processes by mediating particles. Processes mediated by ($\pi$, $\rho$, $a_1$) mesons are displayed on the left, those mediated by the $\omega$ meson on the right.}
  \label{channel_table}
\end{table}

\section{Results}
\begin{figure}
  \begin{center}
  \includegraphics[width=0.91\textwidth]{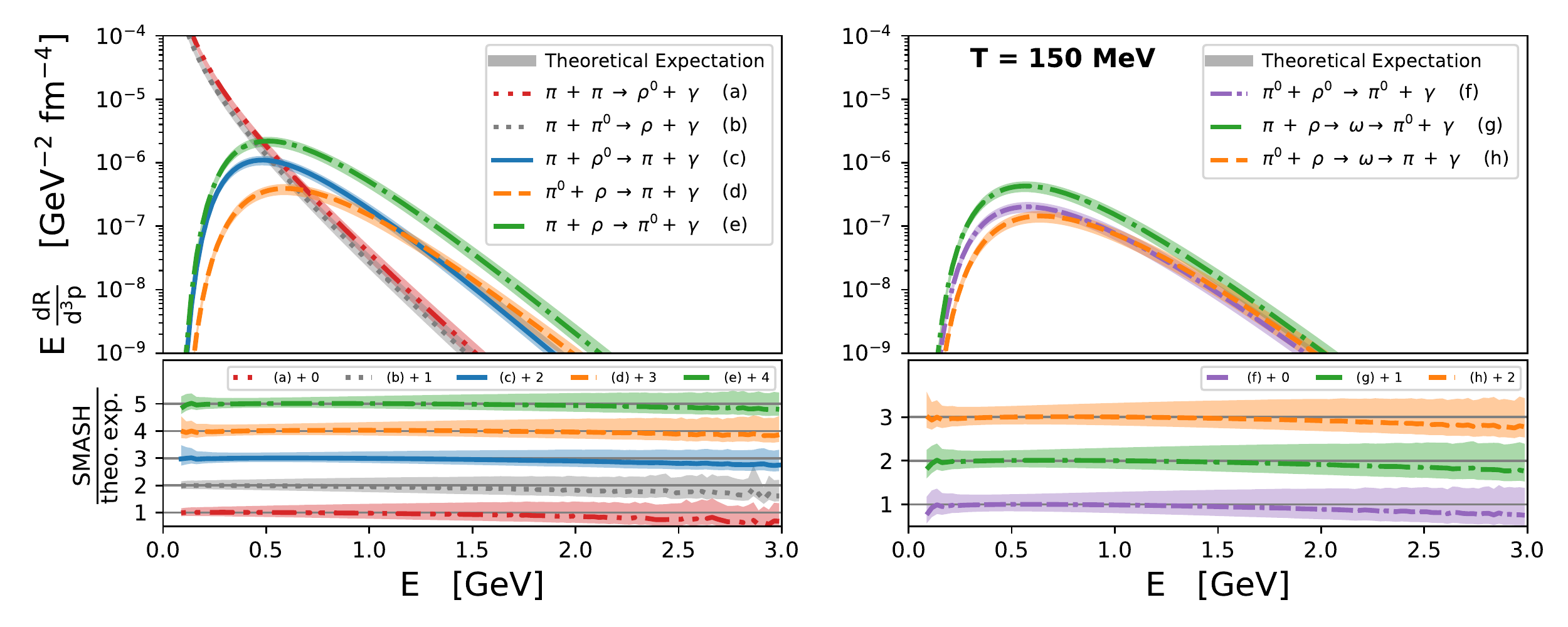}
  \vspace{-0.2cm}
  \caption{Thermal photon rates extracted from SMASH (lines) as compared to semi-analytical theoretical expectations (band) at T = 150 MeV for processes (a)-(g). ($\pi$, $\rho$, $a_1$) mediated processes are displayed on the left, $\omega$ mediated processes on the right. Form factors are not applied and the $\rho$ meson is treated as a stable particle.}
  \label{Theory comparison}
  \end{center}
  \vspace{-0.6cm}
\end{figure}

\begin{figure}[ht]
\begin{minipage}[t]{0.48\textwidth}
    \includegraphics[width=0.95\textwidth]{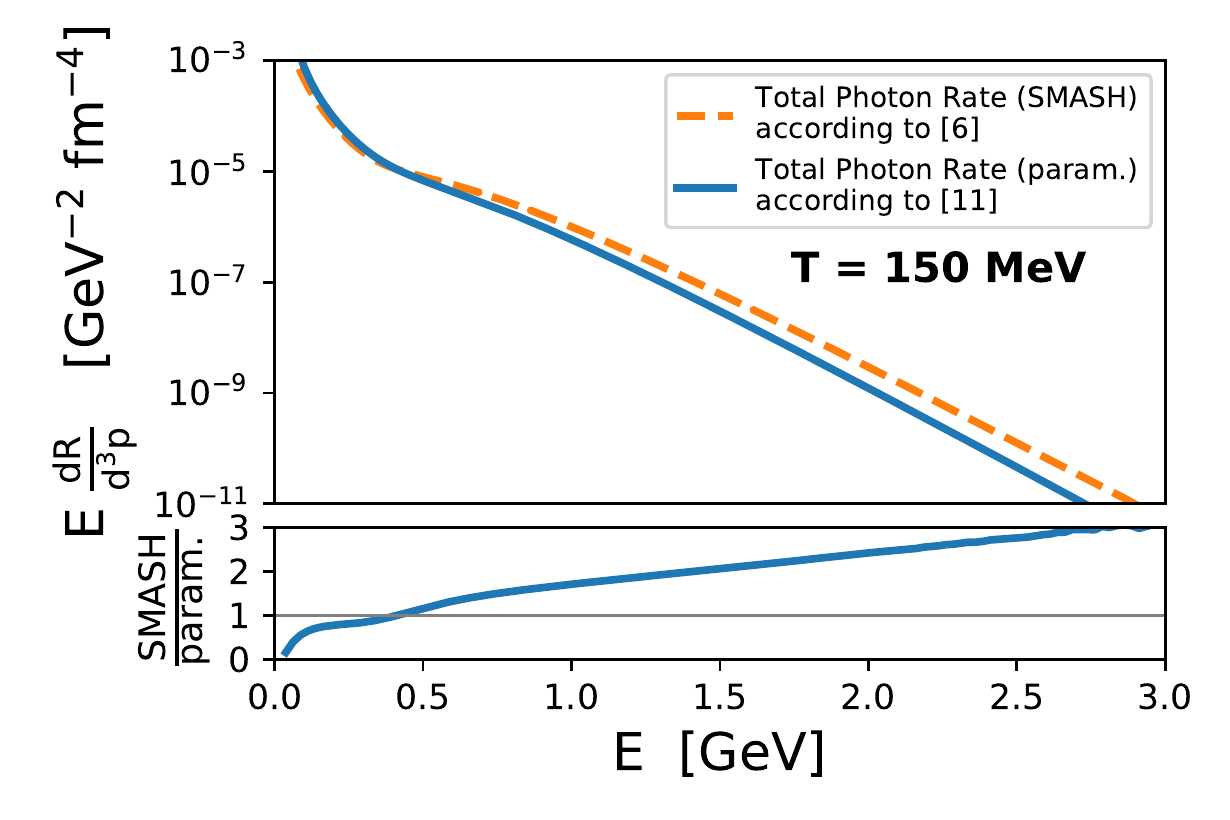}
    \vspace{-0.2cm}
     \caption{Comparison of total parametrized photon rates from \cite{Kapusta:1992gv, Nadeau:1992cn} (blue, solid) to total photon rates from SMASH (orange, dashed) at T = 150 MeV. Form factors are included in SMASH, the parametrizations are correspondingly corrected.}
    \label{Kapusta Comp}
\end{minipage}
\hspace{0.04\textwidth}
\begin{minipage}[t]{0.48\textwidth}
    \includegraphics[width=0.95\textwidth]{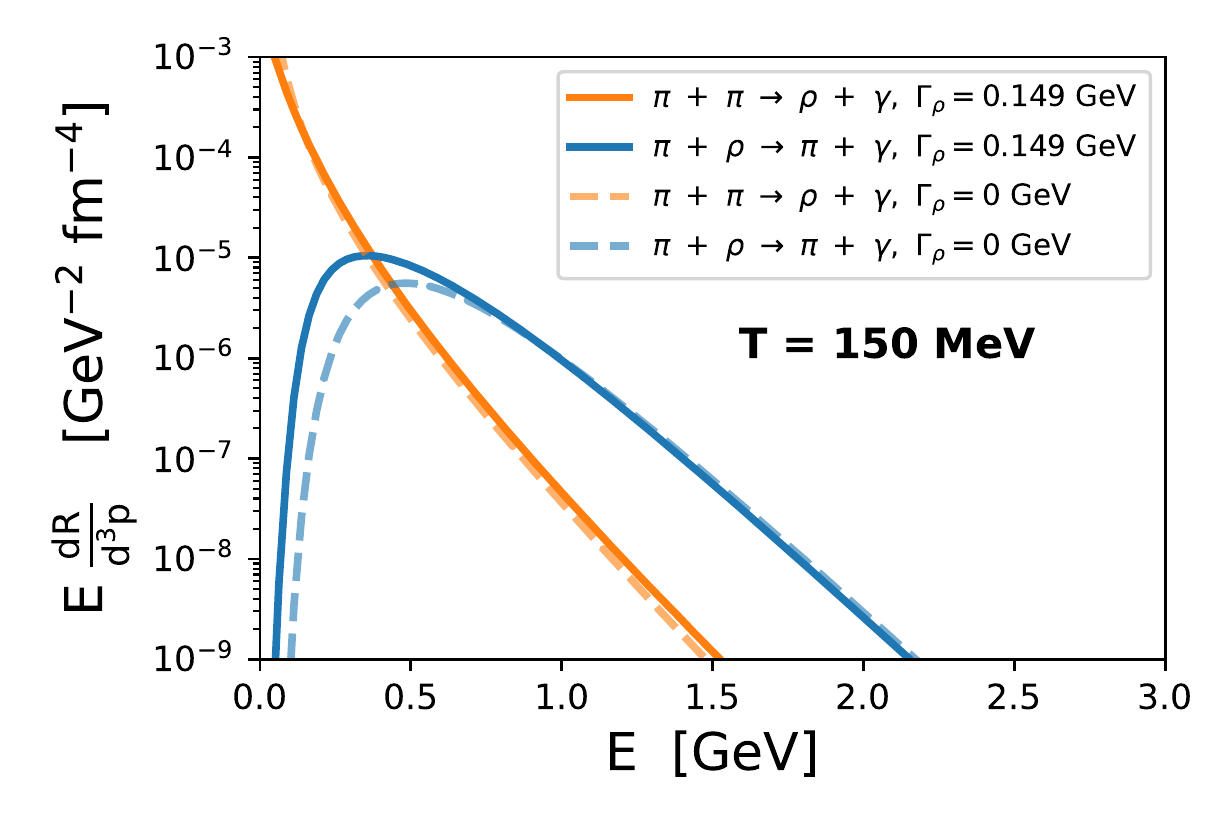}
    \vspace{-0.2cm}
    \caption{Comparison of thermal photon rates for $\Gamma_\rho$ = 0 GeV (dashed) and $\Gamma_\rho$ = 0.149 GeV (solid) for $\pi + \pi \to \rho + \gamma$ (orange) and  $\pi + \rho \to \pi + \gamma$ (blue) processes at T = 150 MeV. Form factors are included.}
    \label{Broad Rho Comp}
\end{minipage}
\end{figure}

The cross sections \cite{phoxtrot} as well as the photon framework are subject to a proof of concept in Figure \ref{Theory comparison}.
The thermal photon rate as obtained from SMASH (lines) is compared to the corresponding semi-analytical expectation values (bands) for each of the processes listed in Table \ref{channel_table}.
The photon rates are extracted in an infinite matter simulation with SMASH at a temperature of 150 MeV. 
Error bands of the theoretical expectation are due to uncertainties in the temperature extraction of the medium.
Within errors, a perfect agreement can be observed between the photon rates resulting from SMASH and those from the semi-analytical benchmark; hence verifying the presented framework. \\
To assess the significance of introducing additional degrees of freedom, a comparison with previous works \cite{Kapusta:1992gv}, as similarly done in \cite{Baeuchle:2009ep}, is performed. In contrast to the above presented framework, the degrees of freedom therein are limited to $\pi$, $\rho$, $\eta$ mesons and the photon, thus lacking processes mediated by $\omega$ and $a_1$ mesons.
In Figure \ref{Kapusta Comp} a comparison of the photon rates obtained from SMASH  (including form factor corrections and having combined processes (e) and (g) as well as (d) and (h)) to parametrizations of the photon rates from \cite{Kapusta:1992gv, Nadeau:1992cn} is shown.
It can be observed that the photon rates determined within \cite{Kapusta:1992gv} yield smaller contributions than SMASH for $E \lesssim 0.4$ GeV and vice versa.
Especially for high photon energies, this discrepancy becomes more pronounced.
It can be deduced that for a realistic description of mesonic photon production, $\omega$ and $a_1$ mediated processes should not be neglected. Similar observations have already been made in \cite{Holt:2015cda}. \\
As mentioned above, the presented framework is capable of describing stable $\rho$ mesons only but can, under certain assumptions, be extended to finite-width $\rho$ mesons.
The effect on the thermal photon rate of treating the $\rho$ meson as a resonance is shown in Figure \ref{Broad Rho Comp}, where the dashed lines correspond to $\Gamma_\rho$ = 0 GeV and the solid lines to $\Gamma_\rho$ = 0.149 GeV.
Only a minor effect is found for $\pi + \pi \to \rho + \gamma$ processes, whereas there is a significant increase of photon production for $\pi + \rho \to \pi + \gamma$ processes especially in the low- and mid-energy regime.
The latter can be explained by the reduced kinematic threshold for the occurrence of such a scattering process since $\rho$ mesons below the pole mass are also accessible once the finite width is taken into consideration.
Whether or not this effect is visible in final particle spectra of simulated heavy-ion collisions is however yet to be assessed as it highly depends on the energy of the produced photons.

\section{Conclusion and Outlook}
Cross sections for the production of photons in hadronic scattering processes have been derived from an effective chiral field theory with mesonic degrees of freedom and implemented into the SMASH transport model.
A proof of concept has been successfully performed by means of a systematic comparison to semi-analytically calculated theoretical expectation values for the thermal photon rate.
Comparisons to previous works describing photon production in hadronic matter have demonstrated the importance of including $\omega$ and $a_1$ mediated production processes.
An extension of the presented framework to finite-width $\rho$ mesons has further shown a significant increase of photon production in $\pi + \rho \to \pi + \gamma$ scattering processes in the low- and intermediate-energy range. \\
This work is a first step to consistently treat photon emission in hybrid hydrodynamics+transport approaches. The 3+1D viscous hydrodynamics code MUSIC \cite{Schenke:2010nt, Ryu:2015vwa} and SMASH \cite{Weil:2016zrk, SMASH-DOI} are particularly well suited for such an approach as both rely on the same underlying field theory for photon production in the hadronic phase. Not only can it be applied to describe a heavy-ion collision at RHIC/LHC energies, but also to benchmark the importance of a non-equilibrium treatment in the late stages of relativistic heavy-ion collisions as compared to a pure hydrodynamic evolution. To this end, the incorporation of bremsstrahlung contributions is essential and constitutes a next step. Such a hybrid model provides a great opportunity to identify and study non-equilibrium effects in the afterburner and to assess the importance of the late stages to contribute to the resolution of the \textit{direct photon flow puzzle} \cite{Gale:2018ofa}.
\\

\noindent \textbf{Acknowledgements}
This project was supported by the DAAD funded by BMBF with Project-ID 57314610.
A.S. acknowledges support by the Stiftung Polytechnische Gesellschaft Frankfurt am Main.
C.G. is supported in part by the Natural Sciences and Engineering Research Council of Canada.
H.E. and J.M.T.R. acknowledge support by the Deutsche Forschungsgemeinschaft (DFG) through the Grant No. CRC-TR 211 “Strong-interaction matter under extreme conditions”.
J.M.T.R. was further supported by the DFG through Grant No. 411563442.
Computational resources have been provided by the Center for Scientific Computing (CSC) at the Goethe- University of Frankfurt and the GreenCube at GSI.








\end{document}